\newif\ifsubmode
\submodefalse

\ifsubmode
  \documentclass[12pt,preprint]{aastex}  
  \received{}
  \revised{}
  \accepted{}

\else
  \documentclass{emulateapj}  
  \slugcomment{Submitted to \apj}
  \usepackage{lscape}
\fi

\shortauthors{Fassnacht et al.}
\shorttitle{Three lenses for the price of one}

\newcommand{\lensa}{``Fred''}
\newcommand{\lensb}{``Ginger''}

\newcommand{\kms}{km\ s$^{-1}$}
\newcommand{\kmsmpc}{km\ s$^{-1}$\ Mpc$^{-1}$}

\begin{document}

\title{Three Gravitational Lenses for the Price of One:\\
 Enhanced Strong Lensing through Galaxy Clustering
\footnote{\rm Based in
part on observations made with the NASA/ESA Hubble Space Telescope,
obtained at the Space Telescope Science Institute, which is operated
by the Association of Universities for Research in Astronomy, Inc.,
under NASA contract NAS 5-26555. These observations are associated
with program \#GO-10158.}
}

 \author{C. D. Fassnacht\altaffilmark{1},
 J. P. McKean\altaffilmark{1}, L. V. E. Koopmans\altaffilmark{2},
 T. Treu\altaffilmark{3}, R. D. Blandford\altaffilmark{4},\\
 M. W. Auger\altaffilmark{1}, T. E. Jeltema\altaffilmark{5}, 
 L. M. Lubin\altaffilmark{1}, V. E. Margoniner\altaffilmark{1}, 
 D. Wittman\altaffilmark{1}}

\altaffiltext{1}{Department of Physics, 
  University of California, Davis, 
  1 Shields Avenue, 
  Davis, CA 95616\\ 
  {\tt fassnacht@physics.ucdavis.edu
}}
 
\altaffiltext{2}{Kapteyn Astronomical Institute, 
    University of Groningen P.O. Box 800, 9700 AV Groningen,
    Netherlands 
}
 
\altaffiltext{3}{Department of Physics, University of California, 
    Santa Barbara 
}
 
\altaffiltext{4}{Kavli Institute for Particle Astrophysics and
    Cosmology, Stanford Linear Accelerator Center, MS 75, 2575 Sand
    Hill Road, Menlo Park, CA 94025
}
 
\altaffiltext{5}{Observatories of the Carnegie Institute of
    Washington, 813 Santa Barbara St., Pasadena, CA 91101 
}

\begin{abstract}
We report the serendipitous discovery of two strong gravitational lens
candidates (ACS J160919+6532 and ACS J160910+6532) in deep images
obtained with the Advanced Camera for Surveys on the {\em Hubble Space
Telescope}, each less than 40\arcsec\ from the previously known
gravitational lens system CLASS B1608+656. The redshifts of both lens
galaxies have been measured with Keck and Gemini:\ one is a member of
a small galaxy group at $z$\,$\approx$\,0.63, which also includes the
lensing galaxy in the B1608+656 system, and the second is a member of
a foreground group at $z$\,$\approx$\,0.43. By measuring the effective
radii and surface brightnesses of the two lens galaxies, we infer
their velocity dispersions based on the passively evolving Fundamental
Plane (FP) relation. Elliptical isothermal lens mass models are able
to explain their image configurations within the lens hypothesis, with
a velocity dispersion compatible with that estimated from the FP for a
reasonable source-redshift range. Based on the large number of massive
early-type galaxies in the field and the number-density of faint blue
galaxies, the presence of two additional lens systems around
CLASS~B1608+656 is not unlikely in hindsight.  Gravitational lens
galaxies are predominantly early-type galaxies, which are clustered,
and the lensed quasar host galaxies are also clustered.  Therefore,
obtaining deep high-resolution images of the fields around known
strong lens systems is an excellent method of enhancing the
probability of finding additional strong gravitational lens systems.
\end{abstract}


\section{Introduction}

Strong gravitational lenses are excellent tools for cosmological and
astrophysical studies \citep[see, e.g.,][]{cskreview}.  Because of
their utility, an increase in the number of lenses is a highly
desirable goal, especially when the numbers become large enough to
marginalize over hidden parameters associated with any given lens
system. The majority of lenses discovered in the last decade were
found through dedicated surveys that used a variety of techniques to
find strong lenses, such as targeting the potential lensed objects
\citep[e.g.,][]{hstlens,class1,class2,winnsearch,wisotzki,pindorlens},
searching around the potential lensing galaxies
\citep[e.g.,][]{goodslens}, or looking for multiple redshifts
associated with a single object in large spectroscopic surveys
\citep[e.g.,][]{bolton1,bolton2,slacs1}.  However, a 
large number of lens systems were discovered serendipitously. In this
paper we report just such a discovery.  We have found two additional
strong lens candidates in a single image centered on a known
gravitational lens, the data of which were obtained with the Advanced
Camera for Surveys
\citep[ACS;][]{acs1,acs2} on the {\em Hubble Space Telescope} (HST).

\begin{figure*} 
  \plottwo{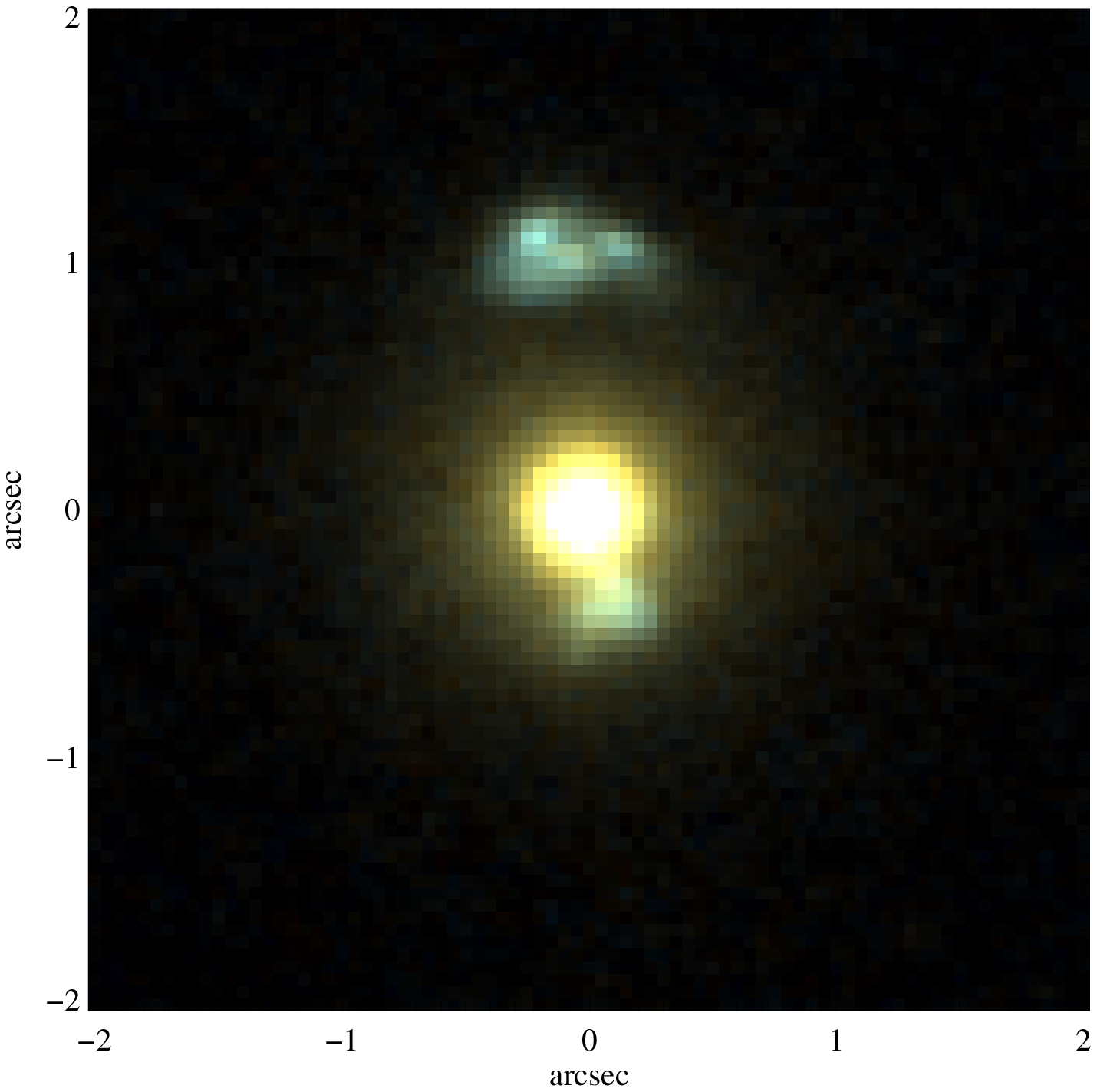}{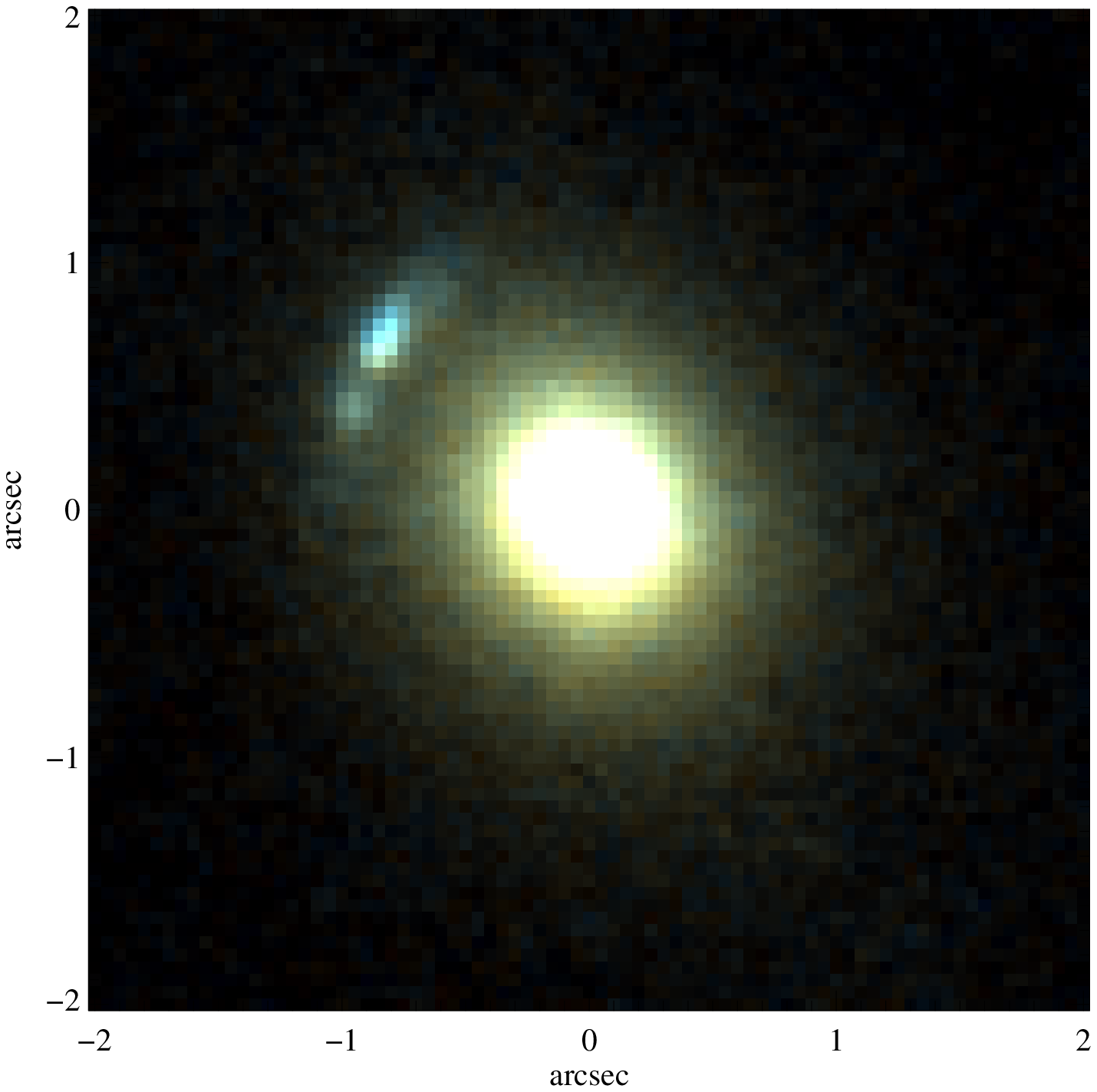}
  \caption{Three-color images of \lensa\ (left) and \lensb\
  (right). The images were constructed from the ACS images, with the
  F606W data in the blue channel, the sum of F606W and F814W in the
  green channel, and F814W in the red channel, respectively.  In each
  image, north is up and east is to the left.  The blue blobs and arc
  stand out clearly from the red lensing galaxies.
  \label{fig_color}}
\end{figure*}

The main goal of the ACS observations was to obtain a high
sensitivity image of the Einstein ring of a previously known
gravitational lens, CLASS B1608+656 \citep{stm1608}.  The Einstein
ring provides further constraints to the lensing mass model and
reduces the uncertainties in the determination of $H_0$ from this
time-delay system \citep[e.g.,][]{Ko01, 1608H0}. The exquisite angular
resolution and surface brightness sensitivity of the ACS imaging also
allows the properties of the B1608+656 field to be studied. In
particular, the contribution to the B1608+656 image splitting by
a small group of galaxies associated with the main lensing galaxy has
been investigated \citep{1608groupdisc}, and a weak lensing analysis of
the mass distribution along the line of sight to B1608+656 is
being conducted.  The deep ACS imaging has also allowed us to find
two additional galaxy-scale gravitational lens systems in the same 
field of view.

In Section 2 we present the ACS and ground-based imaging of the two
lens candidates. The spectra of the lensing galaxies and one of the
lensed sources are presented in Section 3.  We test lensing mass
models for the two systems using a non-parametric lensing code in
Section 4. Finally, in Section 5 we discuss the lensing hypothesis for
both systems and estimate the likelihood of finding two additional
lenses in this field. Throughout this paper we assume a flat Universe
with $\Omega_M = 0.3$, $\Omega_\Lambda = 0.7$, and, unless otherwise
stated, we will express the Hubble Constant as H$_0 = 100~h$~\kmsmpc.

\section{Optical Imaging}

In this section, we present space- and ground-based multi-color optical
imaging of the two lens candidates in the field of B1608+656.

\subsection{Hubble Space Telescope}

High resolution optical imaging of the B1608+656 field was obtained
with the ACS (GO-10158; PI: Fassnacht). The data were acquired over
the course of five visits between 2004 August 24 and September 17. The
Wide Field Channel (WFC) of the ACS was used throughout, providing a
field of view (FOV) of $202\arcsec \times 202\arcsec$ and a scale of
0.05~arcsec~pixel$^{-1}$. Our observations consisted of nine orbits
that used the F606W filter and eleven orbits with the F814W filter,
corresponding to total exposure times of 22516~s and 28144~s,
respectively. The data were reduced in the standard manner using the
{\it stsdas} package within {\sc iraf}\footnote{IRAF (Image Reduction
and Analysis Facility) is distributed by the National Optical
Astronomy Observatories, which are operated by the Association of
Universities for Research in Astronomy under cooperative agreement
with the National Science Foundation.}. The final combined images were
produced using {\it multidrizzle} \citep{multidrizzle}, which also
corrected the data for the ACS geometric distortion.  The area covered
by the final combined image in each filter, defined as the region in
which the weight file had pixel values greater than 2000, is 11.9
arcmin$^2$.  Catalogs of objects in the ACS images were generated
by running SExtractor
\citep{sextractor} with the parameters suggested by
\citet{faintacsgals}. The count rates in the images were converted to
Vega-based magnitudes using the zero points on the ACS web site\footnote{See
\url{http://www.stsci.edu/hst/acs/analysis/zeropoints}}.  Full details
of the acquisition and reduction of our ACS imaging will be presented
in a future paper. A summary of the imaging observations is given in Table
\ref{tab_obsdata}.

\begin{deluxetable}{ccccr}
\tabletypesize{\scriptsize}
\tablecolumns{10}
\tablewidth{0pc}
\tablecaption{Imaging Observations}
\tablehead{
\colhead{}
 & \colhead{}
 & \colhead{}
 & \colhead{}
 & \colhead{$t_{exp}$}\\
\colhead{Date}
 & \colhead{Telescope}
 & \colhead{Instrument}
 & \colhead{Filter}
 & \colhead{(sec)}
}
\startdata
2000 Apr      & P60        & CCD13   & $g$ & 7200 \\
2000 Apr      & P60        & CCD13   & $r$ & 3600 \\
2000 Apr      & P60        & CCD13   & $i$ & 3000 \\
2000 Jul      & P60        & CCD13   & $g$ & 5400 \\
2004 Aug/Sep  & HST & ACS/WFC & F606W & 22516 \\
2004 Aug/Sep  & HST & ACS/WFC & F814W & 28144 \\
\enddata
\label{tab_obsdata}
\end{deluxetable}

\begin{figure}[!]
  \plotone{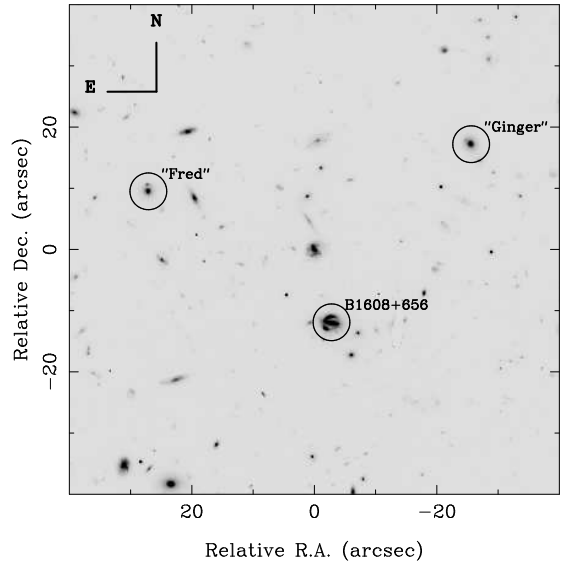} 
  \caption{Wider
  field of view, showing B1608+656 and the two additional strong lens
  candidates in relation to it. The image was obtained through the
  F814W filter with the ACS. 
   \label{fig_fc}} 
\end{figure}

A visual inspection of the ACS images, undertaken with the goal of
evaluating the properties of the galaxies surrounding B1608+656,
revealed two objects with lens-like morphologies.  Each consists of a
reddish early-type galaxy with a nearby blue arc or multiple blue
blobs, similar to other lenses found in HST imaging
\citep[e.g.,][]{ratnatunga,goodslens,blakeslee,slacs1}.  The two
lens candidates, ACS J160919+6532 and ACS J160910+6532 are shown in
Figure~\ref{fig_color}.  For simplicity, the two lens candidates will
hereafter referred to as \lensa and \lensb, respectively.  In
Fig.~\ref{fig_fc} we show a larger field of view that includes the two
lensing candidates and B1608+656.  Both lens candidates are in close
proximity to B1608+656 on the sky:
\lensa\ lies $\sim 36$\arcsec\ to the northeast, whereas
\lensb\ is $\sim 37$\arcsec\ to the north-northwest.  The 
coordinates of the lens candidates are given in Table
\ref{tab_lenscoords}. 

\begin{deluxetable}{lccc}
\tabletypesize{\scriptsize}
\tablecolumns{4}
\tablewidth{0pc}
\tablecaption{Lens System Coordinates}
\tablehead{
\colhead{Name}
 & \colhead{RA (J2000)}
 & \colhead{Dec (J2000)}
 & \colhead{$z_{l}$}
}
\startdata
\lensa\ & 16 09 18.760 & $+$65 32 49.72 & 0.6321 \\
\lensb\ & 16 09 10.292 & $+$65 32 57.38 & 0.4264 \\
\enddata
\label{tab_lenscoords}
\end{deluxetable}

\lensa\ consists of a red spheroidal galaxy with two blue candidate
lensed images to the north and south. The blue image to the north
 is extended in an east/west direction, while the southern
blue image is both fainter and covers a smaller angular
size on the sky. The sizes and surface brightnesses of
the two blue images are consistent with gravitational lensing. Their
separation is $\sim$1\farcs5.

The second lens candidate, \lensb, is a bright red elliptical with an
extended blue gravitational arc to the north east. The narrow arc-like
feature appears to curve toward the red galaxy, which is consistent
with lensing. There is also evidence of substructure in the arc, which
is presumably due to clumps of star formation. No obvious counter-image
can visually be identified in the color map. However, the
galaxy-subtracted residuals (Section~4) do show a faint possible
counter-image.

Fitting de Vaucouleurs profiles to the lens galaxies -- after masking
the lensed features -- yields the parameters listed in
Table~\ref{tab_lensprops} \citep[for a description of the procedure
see, e.g.,][]{T06}.  Observed quantities are transformed into rest
frame quantities as described in \citet{T01}, yielding effective radii
and surface brightnesses of R$_{\rm e}$=$4.76\pm0.48$ and
$3.60\pm0.36$ kpc and SB$_{\rm e}$=$20.42\pm0.05$ and $20.81\pm0.05$
mag arcsec$^{-2}$ for \lensa and \lensb, respectively.  These values
were calculated assuming that $h = 0.7$. Correcting for evolution to
$z=0$ adopting $d\log (M/L_{\rm
B})/dz=-0.72\ \pm0.04$ \citep{T05}, and assuming that the lens
galaxies obey the Fundamental Plane \citep[FP;][]{D87,DD87}
relationship yields estimates for the central velocity dispersions of
$\sigma_{\rm FP}$=$180\pm51$ and $142\pm32$ \kms\ for \lensa
and \lensb, respectively. The uncertainty on $\sigma_{\rm FP}$ is
dominated by the uncertainty in the evolutionary correction and in the
intrinsic thickness of the FP (assumed to be 0.08 in $\log {\rm
R}_{\rm e}$).

\subsection{Palomar 60-inch Telescope}

Ground-based imaging of the B1608+656 field was obtained as part of a
program to investigate the environments of strong gravitational lenses
\citep[e.g.,][]{1608groupdisc}.  The observations were conducted
using the Palomar 60-Inch Mayer telescope (P60), with the CCD13
detector.  The CCD provides a FOV of approximately
13\arcmin$\times$13\arcmin, with a pixel size of 0\farcs379. The data
were taken using the Gunn $g$, $r$, and $i$ filters in 2000 April.
The seeing was approximately 1\farcs2 through the $r$ and $i$ filters,
while it was 1\farcs5 through the $g$ filter.  Exposure times are
given in Table~\ref{tab_obsdata}.  A further observing session in 2000
July was used to increase the depth of the $g$ band data.  In these
observations, the seeing was $\sim$1\farcs2.  The conditions during
both observing sessions were photometric.  The data were reduced using
standard {\sc iraf} tasks, and then the $g$-band data from the two observing
sessions were combined.  Astrometric solutions were computed with the
aid of positions obtained from the USNO-A2.0 catalog \citep{usno} while
photometric solutions were derived from observations of several Gunn
standard stars \citep{gunnstd}.  Given the seeing and the faintness of the
background objects, the background sources are not resolved in the
P60 imaging.  Therefore only total (lens + source) magnitudes are
given in Table~\ref{tab_lensprops}.

\ifsubmode
   \begin{deluxetable}{lccccccccc}
\else
   \begin{deluxetable*}{lccccccccc}
\fi
\tabletypesize{\scriptsize}
\tablecolumns{10}
\tablewidth{0pc}
\tablecaption{Lens System Photometric Properties}
\tablehead{
\colhead{}
 & \colhead{}
 & \colhead{}
 & \colhead{}
 & \colhead{}
 & \colhead{}
 & \colhead{R$_{\rm e,F606W}$\tablenotemark{b}}
 & \colhead{R$_{\rm e,F814W}$\tablenotemark{b}}
 & \colhead{}
 & \colhead{}\\
\colhead{Name}
 & \colhead{$g_{tot}$\tablenotemark{a}}
 & \colhead{$r_{tot}$\tablenotemark{a}}
 & \colhead{$i_{tot}$\tablenotemark{a}}
 & \colhead{F606W$_{\rm l}$\tablenotemark{b}}
 & \colhead{F814W$_{\rm l}$\tablenotemark{b}}
 & \colhead{(\arcsec)}
 & \colhead{(\arcsec)}
 & \colhead{F606W$_{\rm s}$\tablenotemark{c}}
 & \colhead{F814W$_{\rm s}$\tablenotemark{c}}
}
\startdata
\lensa\ & 22.5 & 21.6 & 21.0 & 21.9 & 20.3 & 0.62 & 0.69 & 23.9 & 23.1 \\
\lensb\ & 22.1 & 20.8 & 20.4 & 21.1 & 20.0 & 0.64 & 0.66 & 24.6 & 24.0 \\
\enddata
\label{tab_lensprops}
\tablenotetext{a}{Magnitudes are the ``MAG\_AUTO'' magnitudes returned by
SExtractor, with the default values for the Kron factor and minimum radius.}
\tablenotetext{b}{Magnitudes and effective radii of the lensing galaxy
from the best fit 
de Vaucouleurs model; not corrected for Galactic extinction}
\tablenotetext{c}{Brighter of the two lensed images.}
\ifsubmode
   \end{deluxetable}
\else
   \end{deluxetable*}
\fi

\section{Spectroscopy \& Redshifts}

Several attempts to obtain the redshifts of the lenses and background
sources have been made.  A summary of the spectroscopic observations
is given in Table~\ref{tab_specdata}.  The first observation, with the
Echellete Spectrograph and Imager \citep[ESI;][]{esi} on the
W. M. Keck II Telescope, was obtained as a part of the general
investigation of the environment of the B1608+656 lens system.  The
subsequent observations, with the Low Resolution Imaging Spectrograph
\citep[LRIS;][]{lris} at Keck and the Gemini Multi-object Spectrograph
\citep[GMOS;][]{gmos} on Gemini-North, were targeted specifically at the lens
candidates.  The LRIS observations were obtained on subsequent nights
and consisted of one exposure on each lens candidate.  Each
observation was taken at the very end of the night, ending after 18
degree twilight, so the blue-side observations were swamped by the sky
emission.  In contrast, the GMOS observations of \lensa\ were obtained
in excellent conditions, with a dark sky and seeing of $\sim$0\farcs6.
The three GMOS exposures of the system were dithered in the spectral
direction in order to fill in the gaps between the chips.
The LRIS and ESI data were reduced using scripts that provided
interfaces to the standard {\sc iraf} tasks.  In addition, a portion of the
ESI data reduction was performed using custom IDL scripts.  The GMOS
data were reduced using the {\it gmos} package in {\sc iraf}. The
wavelength calibration for all of the exposures was based on arclamp
spectra obtained adjacent in time to the science observations.

\begin{figure}
  \plotone{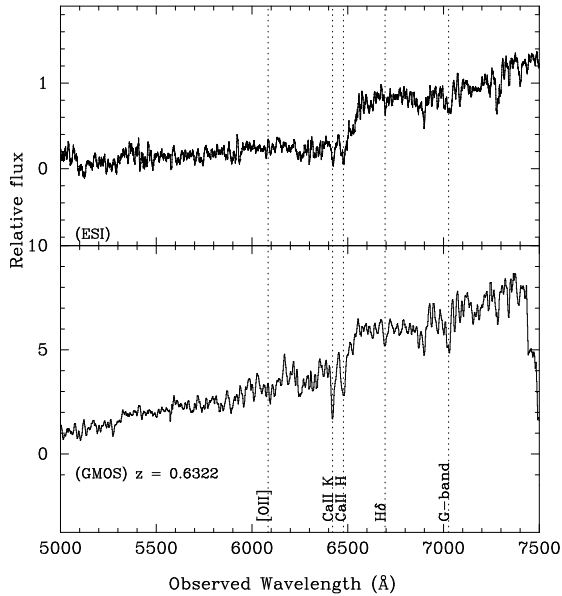}
  \caption{
   Optical spectra of the \lensa\ lens candidate, obtained with Keck/ESI
   (top) and Gemini/GMOS (bottom).  The ESI spectrum has been smoothed
   with a boxcar of width 8.3\AA\ (35~pixels), while the GMOS spectrum
   has been smoothed with a boxcar of width 9.2\AA\ (5~pixels).
   The observed absorption features, as well as the expected location
   of the [\ion{O}{2}] emission, have been marked.
   \label{fig_spec6}}
\end{figure}

\begin{figure}
  \plotone{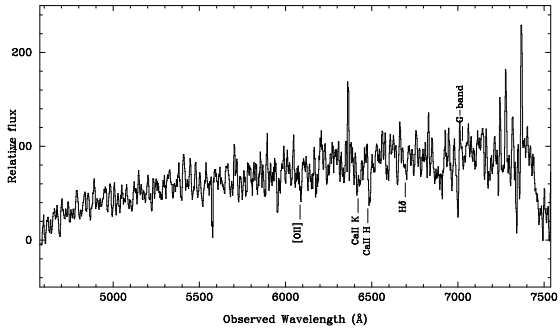}
  \caption{
  Optical spectrum of the \lensa\ background object, obtained with
  GMOS.  All of the marked features are associated with the lensing
  galaxy.  The emission lines correspond to regions in which bright
  night-sky lines have been imperfectly subtracted.  No clear emission
  lines from the background source are seen.
  \label{fig_spec6_src}} 
\end{figure}

The first lens candidate, \lensa\, has a redshift of $z_\ell =
0.6321$, based on multiple absorption features (\ion{Ca}{2},
H$\delta$, and the G-band) seen in the ESI and GMOS spectra
(Figure~\ref{fig_spec6}). The \lensa\ redshift places it in a small
group that also includes the lensing galaxy in the B1608+656 system
\citep{1608groupdisc}. The ESI observations were obtained before
the system was identified as a lens candidate.  Although the position
angle (PA) of the slit was approximately correct for the system
morphology, no trace of the lensed source is seen in the spectrum.  On
the other hand, the GMOS spectrum of this object was obtained with a
slit PA chosen specifically to cover both the lensing galaxy and the
lensed source.  The brighter of the two blue blobs is clearly seen in
the two-dimensional spectrum and is spatially separated from the lens
galaxy spectrum, allowing a separate spectrum to be extracted for the
lensed image (Figure~\ref{fig_spec6_src}).  However, no clear emission
or absorption features, beyond those due to contamination from the
lens galaxy light and imperfect subtraction of the night sky lines,
are seen in the spectrum of the source.  Given the blue color of the
lensed source, we expect to see emission lines in the spectrum; the
lack of emission features allows us to place tentative limits on the
redshift of the background source.  Emission from [\ion{O}{2}]
$\lambda$3727 shifts out of the range covered by the GMOS spectrum at
a redshift of $z_s = 1.0$, while Ly$\alpha$ emission enters the
spectral range at $z_s = 2.8$.  We therefore assume that the
background object falls within this redshift range.  For completeness
we have to consider the
possibility that the blue knots are star formation associated with the
lensing galaxy.  However, if this were the case, we would expect
strong [\ion{O}{2}] emission at a wavelength of 6083\AA.  We do not
see any evidence for [\ion{O}{2}] emission at 6083\AA\ in any of the \lensa
spectra and therefore conclude that the arc is not due to star
formation in the lensing galaxy.

\begin{figure}
  \plotone{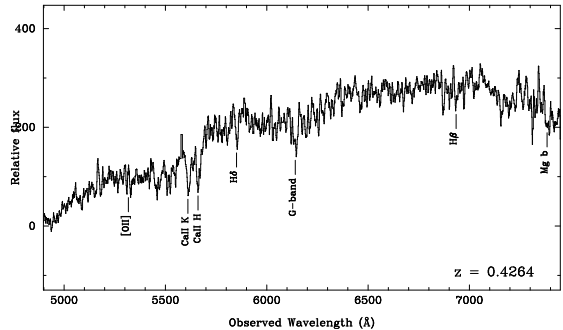}
  \caption{
   The optical spectrum of the \lensb\ lens candidate taken with LRIS
   on the Keck Telescope. 
   The data have been smoothed with a boxcar of width 5
   pixels, corresponding to 6.25~\AA. The spectrum is dominated by the 
   emission from the lens, whose spectral shape is consistent with an
   early-type galaxy.  The redshift of the lens is $z=0.4264$.  There
   is no evidence of ongoing star formation (the expected position of
   the [\ion{O}{2}] emission line is marked). 
   \label{fig_spec4}}
\end{figure}

The other lens candidate, \lensb, has a spectrum typical of an
early-type galaxy (Figure~\ref{fig_spec4}).  A number of absorption
features, including lines due to \ion{Ca}{2} H and K, H$\delta$, and
the G-band give a redshift of $z_{lens} = 0.4264$.  Although the slit
PA was chosen to cover both the galaxy and the brightest portion of
the arc, no redshift for the background source was obtained.  This may
just be due to the short integration time and bright night sky.  The
redshift of the lens galaxy places it in another group detected along
the line of sight to B1608+656.  This group has a mean redshift of $z
= 0.426$ \citep{1608groupdisc}.

\ifsubmode
   \begin{deluxetable}{lcccccrrrc}
\else
   \begin{deluxetable*}{lcccccrrrc}
\fi
\tabletypesize{\scriptsize}
\tablecolumns{10}
\tablewidth{0pc}
\tablecaption{Spectroscopic Observations}
\tablehead{
\colhead{}
 & \colhead{}
 & \colhead{}
 & \colhead{}
 & \colhead{}
 & \colhead{Wavelength}
 & \colhead{Slit Width}
 & \colhead{Slit}
 & \colhead{$t_{exp}$}
 & \colhead{}
\\
\colhead{Date}
 & \colhead{Target}
 & \colhead{Telescope}
 & \colhead{Instrument}
 & \colhead{Grating}
 & \colhead{Coverage}
 & \colhead{(\arcsec)}
 & \colhead{PA}
 & \colhead{(sec)}
 & \colhead{SNR\tablenotemark{a}}
}
\startdata
2001 Jul 23 & \lensa\ & Keck II  & ESI    & 175\tablenotemark{b}
 & 3887--10741 & 1.0 & +11.8    & 3600 & 2.5 \\
2005 Apr 12 & \lensb\ & Keck I   & LRIS/R & 600/5000            
 & 4899--7488 & 1.0 & +48.2 & 1800 & 8 \\
2005 Apr 13 & \lensa\ & Keck I   & LRIS/R & 600/5000            
 & 4908--7452 & 1.0 & +8.2  & 1800 & 8 \\
2005 May 13 & \lensa\ & Gemini-N & GMOS   & B600                
 & 4587--7478 & 1.0 & +8.2  & 4500 & 12 \\
\enddata
\tablenotetext{a}{Value given is the average SNR pix$^{-1}$ in the
region just redward of the \ion{Ca}{2} absorption lines.}
\tablenotetext{b}{Cross-dispersed.}
\label{tab_specdata}
\ifsubmode
   \end{deluxetable}
\else
   \end{deluxetable*}
\fi

\section{Gravitational Lens Models}

\begin{figure*} 
  \plottwo{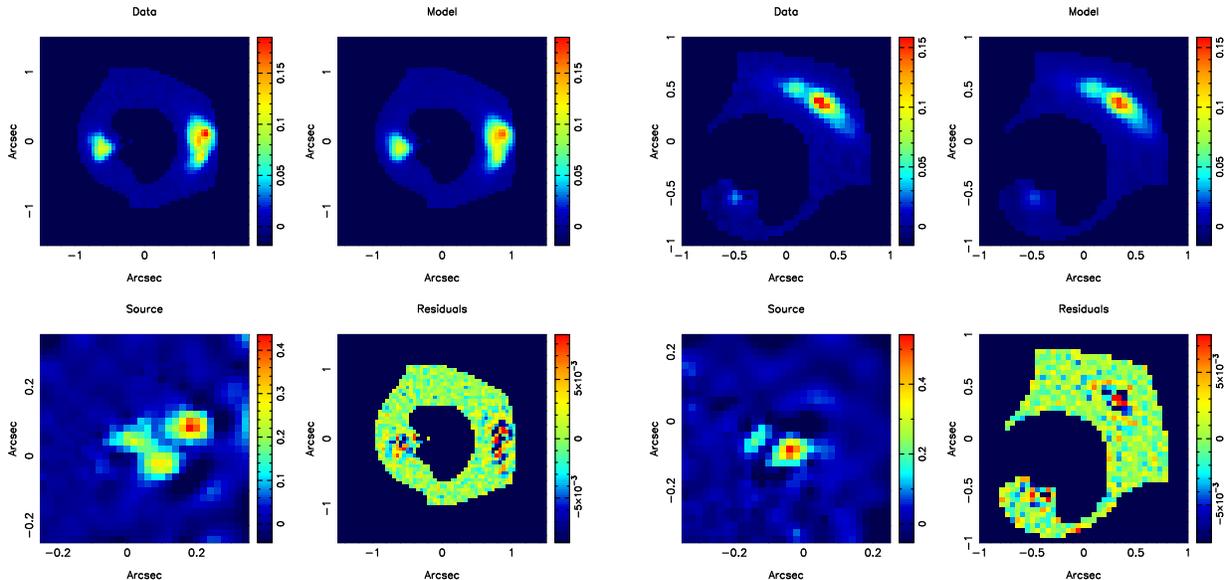}{fig6b.eps} 
  \caption{{\bf
  (Left)} Model results for the \lensa\ lens candidate.  Panels show
  observed data with lensing galaxy subtracted (top left), the
  reconstructed lensed image (top right), the reconstructed source
  (bottom left), and the residuals between observed and model lensed
  emission (bottom right).  The panels are rotated by $-93.8^\circ$,
  so that the approximate orientation is east up and north to the
  right.
  {\bf (Right)} Idem, but for the \lensb\ lens candidate.  
  \label{fig_mod6}} 
\end{figure*}

We reconstruct the lensed images and source of both systems to test the
lens hypothesis and assess whether a relatively simple mass model can
explain the observed lens-like features in Figure~1. We adopt a
Singular Isothermal Ellipsoid (SIE) lens mass model \citep{siemodel},
which describes the mass distribution (stellar plus dark
matter) in the inner regions of massive lens-galaxies extremely well
\citep[e.g.,][]{tk_galevol,slacs3}.  In addition, we add where necessary 
an external shear to account for the group environment
\citep{1608groupdisc}.  The modeling is done via a
non-parametric reconstruction method, which is described in more
detail in \citet{tk_galevol} and \citet{lvek04}. We regularize the
solutions somewhat to attain a smoother solution; see
\citet{lvek04} for a proper discussion.
The modeling is conducted on the the F606W ACS images, after subtracting
off the emission from the lensing galaxy.  The galaxy-subtracted images
are shown in the upper left panels in Figure~\ref{fig_mod6}.  The
\lensa\ image shows clearly the two bright lensed images.  In the
\lensb\ image, the arc is seen in the top right corner of the 
galaxy-subtracted image, while there is a hint of a counter-image in
the lower left corner

The resulting non-parametric source and lens models are shown in
Figure~\ref{fig_mod6} for both systems.  The resulting SIE and
external shear parameters, for the mass models centered on the
brightness peaks of the observed galaxies, are given in
Table~\ref{tab_models}.  Note that the PA values are given in the
frame of the image and are rotated $-$93.8$^\circ$ with respect to the
PA on the sky. The bracketed quantities are fixed at the observed
values.

\begin{deluxetable}{ccc}
\tabletypesize{\scriptsize}
\tablecolumns{10}
\tablewidth{0pc}
\tablecaption{Lens Model Parameters}
\tablehead{
\colhead{Parameter}
 & \colhead{\lensa}
 & \colhead{\lensb}
}
\startdata
$b_{\rm SIE}$      & 0\farcs73  & 0\farcs61 \\
$\theta_{\rm SIE}$ & \nodata    & [$-49^\circ$]\tablenotemark{a} \\
$q_{\rm SIE}$      & [1.0]\tablenotemark{a}      & 0.84 \\
$\gamma_{\rm ext}$ & 0.056      & \nodata \\
$\theta_{\rm ext}$ & 131$^\circ$ & \nodata \\
\enddata
\tablenotetext{a}{Held fixed at observed value.}
\label{tab_models}
\end{deluxetable}

\section{Discussion\label{sec_disc}}

The first question to address is whether or not the candidates
presented in this paper are actually gravitational lenses.  Of course,
a measurement of the redshifts of the background objects would make
the lens hypothesis more secure.  The current set of spectra were
obtained with limited observing time or wavelength coverage, or were
observed with the slit at a non-optimal PA.  Therefore, a
more dedicated observing campaign may yet yield the redshifts,
especially at shorter wavelengths where the contrast between the lens
and the source is improved \citep[as for HST1543, see][]{tk_galevol}.
The blue colors of the background objects suggest that their spectra
may contain emission lines and thus that it may be possible to obtain
redshifts in spite of the faintness of the objects.  Even without
redshifts, however, the system morphologies and the surface
brightnesses of the background objects are consistent with
gravitational lensing, as can be seen from the lens modeling.

\subsection{The lens model for \lensa}

Based on several arguments -- besides the similar colors and typical
lens-geometry of the two galaxy-subtracted residual images -- we
strongly believe that \lensa\ is a {\em second} strong-lens system in
the field of the B1608+656 lens system. (1) The source brightness
distribution of \lensa\ shows structural features that correspond to
the features seen in the higher magnification image, as expected under
the lens hypothesis. However, when mapped onto the second, much
fainter, image, it also matches the somewhat triangular structure of
that image extremely well, with overall residuals between the observed
images and the lens model at an RMS level of $<10^{-2}$.  There is no
{\em a priori} reason that this mapping onto the second system should
be successful if the system were not a lens system.  (2) The Einstein
radius of $b_{\rm SIE}$=0\farcs73, determined from the best-fit model,
implies an approximate stellar velocity dispersion $\sigma_{\rm SIE}$
between 290 and 210 \kms\ for source redshifts between $z_{\rm s}$=1.0
and 2.0, respectively.  These values are typical for most lens
galaxies (i.e., around L$_*$) and also agree well with the brightness
of the galaxy and absence of emission lines in the spectra. A direct
measurement of its velocity dispersion and source redshift, however,
can secure this agreement more accurately.  (3) The SIE model for
\lensa\ requires an external shear of $\gamma_{\rm ext}\approx 0.06$
with a PA of 225~degrees. This is the direction of the previously
known lens system, B1608+656, (see Figure~2) and could be caused by its
group environment \citep{1608groupdisc}, although the group parameters
are not well constrained by the current data. (4) The last piece of
(circumstantial in this case) evidence, is the good agreement between
the ellipticity of the stellar light and that of the SIE mass model.
This correlation is typically very strong \citep[RMS of $\sim$0.1 in
$q_*/q_{\rm SIE}$; see][]{slacs3} for elliptical-galaxy systems with a
significant stellar mass component inside the Einstein radius.

The morphology of the lensed source is unusual.  However, given its
blue color and presumed (i.e., relatively high) redshift one might
expect the emission from such a galaxy to be dominated by knots of
star formation.  In fact, the source shape is similar to several of
the reconstructed lensed sources in the SDSS Lens ACS Survey (SLACS)
sample of \citet{slacs1}.  Generally speaking, the increased abundance
of peculiar and irregular galaxies at high redshifts and faint
magnitudes is well established from deep
surveys \citep[e.g.,][]{HDFmorph}.  Once again, the facts that the
unusual morphology is seen in both components and that the lens model
so clearly maps the brighter image into the fainter one are strong
arguments in favor of the lensing hypothesis.

The projected mass of the lensing galaxy, within the Einstein ring
radius, is given by
$$
M_E = \frac{c^2}{4 G} \frac{D_\ell D_s}{D_{\ell s}} \theta_E^2
$$
where $\theta_E$ is the Einstein ring radius in angular units and
$D_\ell$, $D_s$, and $D_{\ell s}$ are the angular diameter distances
to the lens, to the background source, and between the lens and the
background source, respectively.  The Einstein ring radius is given
above, and corresponds to half the separation between the two lensed
images. The derived mass of the lensing galaxy is $6.9 \times 10^{10}
(D_s / D_{\ell s}) h^{-1} M_\odot$.  For source redshifts between 1.0
and 2.0, this corresponds to masses between 2.2 and $1.1 \times
10^{11} h^{-1} M_\odot$. Adopting the surface photometry derived in
\S~2.1, this corresponds to a $B$-band mass-to-light ratio inside the 
cylinder of radius equal to the Einstein Radius of 13.6 $h$ (6.8 $h$) in
solar units for $z_{\rm s}=1.0$ (2.0).  These values are consistent
with those found for other lens galaxies \citep[e.g.,][]{tk_galevol},
lending further support to the lensing hypothesis for \lensa.

The velocity dispersion estimated via the FP in \S~2.1 provides a
further consistency check on the lensing interpretation, or an
estimate of the redshift of the background source if the lensing
hypothesis is accepted \citep[e.g.,][]{Ko00}. In fact, empirical
evidence suggests that at scales comparable to the effective radius,
the ratio of stellar to SIE velocity dispersion, $f_{\rm
SIE}=\sigma_*/\sigma_{\rm SIE}$, is close to unity 
\citep[e.g.,][]{Ko00,vdv03,tk_galevol,T06,slacs3}.
The estimated velocity dispersion from the FP ($\sigma_{\rm
FP}=180\pm51$ \kms) is consistent with that of the best fitting SIE
for plausible $z_s$, consistent with the lensing
hypothesis. Conversely, adopting the lens hypothesis, $\sigma_{\rm
FP}$ would imply $z_s>1.52$ for $f_{\rm SIE}=1.0\pm0.1$.

\subsection{The lens model for \lensb}

This system is significantly less constrained because the presumed
counter-image is very faint and the brightness distribution of the
images lacks very distinct structure.  Due to the lack of constraints,
only the lens strength ($b_{\rm SIE}$) and ellipticity ($q_{\rm SIE}$)
are varied in the modeling, and no external shear is used. The
resulting reconstructed source consists of two features that
correspond to those seen in the brighter of the two (lensed) images.
Even though both the geometry and the relative brightnesses of the
images can be fit by a model with several free parameters, the models
are very tentative.  The Einstein radius of $b$=0\farcs61, determined
from the best-fit model, implies an approximate stellar velocity
dispersion between 205 and 180 \kms\ for source redshifts between
$z_{\rm s}$=1.0 and 2.0, respectively. Following the same arguments as
for \lensa we can obtain consistency checks or an estimate for $z_{\rm
s}$, by comparing the results of the lens model with the surface
photometry. The Einstein radius implies a mass within the cylinder of
7.3 (5.4)$\times10^{10}$ $h^{-1}$ M$_{\odot}$ for $z_{\rm s}=1.0$
(2.0). This corresponds to a $B$-band mass-to-light ratio of 11.4 $h$
(8.4 $h$) in solar units for $z_{\rm s}$=1.0 (2.0), which is
consistent with typical values found for other lens galaxies at
similar redshift
\citep{slacs3}. The velocity dispersion implied by the FP,
$\sigma_{\rm FP}=142\pm32$\kms, is somewhat (but not significantly for the
higher source redshift range) smaller than that obtained from the lens
model for $z_s$ in the range 1--2. Thus, under the lens hypothesis
either $z_s>2.2$ is required ($z_s>3$ for $f_{\rm SIE}=1.0\pm0.1$) or
$f_{\rm SIE}<1$, possibly indicating extra convergence from the
environment of \lensb. None of these arguments appear conclusive as to
the lensing nature of \lensb.  Whether \lensb\ is a {\sl third} strong
lens in the field of B1608+656 will probably require a direct
measurement of the source redshift and a more definitive detection
of the counter-image.

\section{Summary \& Conclusions}

Our investigations have shown that the single ACS pointing centered on
B1608+656 contains two additional strong lens candidates.  This result
implies that there are one to two additional lenses in a
$\sim$10~arcmin$^2$ area of the sky, giving {\em a posteriori} lensing
rates of 0.07-0.46 arcmin$^{-2}$ if only \lensa is a real lens, and 
0.14--0.59 arcmin$^{-2}$ if both of the candidates are real lenses 
\citep[68\% limits assuming Poisson statistics;][]{gehrels86}.  
These rates should be contrasted with the results of other HST
lens-search campaigns, which have found lower lensing rates.  For
example, 10 lens candidates, two of which have been confirmed, were
found in the $\sim$600~arcmin$^2$ of the Medium Deep Survey
\citep{ratnatunga}, for a lensing rate of $\leq$0.02 arcmin$^{-2}$.
Also, the search in the Great Observatories Origins Deep Survey
(GOODS) ACS data by
\citet{goodslens} resulted in six lens candidates in $\sim$300
arcmin$^2$, once again giving a lensing rate of at most $\sim$0.02
arcmin$^{-2}$.

In hindsight, it might not be surprising that the lensing rate in the
B1608+656 field is an order of magnitude higher than those in the
larger surveys, although the effect of small number statistics should
not yet be discounted. Qualitatively this is easy to understand,
because the field that is being imaged in these observations is not a
random line of sight.  First, the targeted field is already known to
contain a massive early-type lens galaxy.  These galaxies often are
found in dense environments such as groups and clusters
\citep[e.g.,][]{morphdensity}.  Spectroscopic investigations of
this particular field have revealed the presence of at least three
galaxy groups along the line of sight to the
B1608+656 lens system, including a group that is physically associated
with the lensing galaxy \citep{1608groupdisc}. Second, the lensed source
in the B1608+656 system is itself a massive early-type galaxy
exhibiting AGN activity in the radio domain
\citep{stm1608,zs1608}.   Both as a massive galaxy and as a 
radio source \citep[e.g.,][]{ASradiogroups}, the background source can
be expected to reside in an overdense environment.  The lensed object
in the B1608+656 system is at a redshift of $z_s = 1.394$
\citep{zs1608}.  If the background sources in \lensa\ 
and \lensb\ were at the same redshift, their angular separations
from the B1608+656 system would correspond to $\sim 210
h^{-1}$ kpc, not unusual for a group of galaxies. Third, by obtaining
much deeper than normal space-based imaging of this field, we have
started to pick up the ubiquitous population of faint blue galaxies.
Depending on the steepness of the luminosity function of this
population of galaxies, magnification bias can lead to a higher
lensing rate than expected based on the density of galaxies at the
limiting magnitude of the imaging \citep{slacs1}.

Being quantitative about the expected lensing rate is rather more
difficult because it depends on many unknown factors such as the
redshift and mass distributions of the potential lensing galaxies and
the redshift distribution of the sources, and thus is beyond the scope
of this paper.  However, to test the plausibility of the qualitative
argument, we take a simple path to estimate the expected lensing rate
in a deep ACS image such as this one.  Our assumptions are that (1) only
early-type galaxies will act as lensing galaxies and (2)
all of the lenses will have lensing cross-sections of $\sim$1 arcsec$^2$.
The lensing cross-section is estimated by taking a circular
area with a diameter equal to the typical image
separation for the lenses discovered by the Cosmic Lens All-Sky Survey
\citep{class2}.  We define the early-type galaxies as luminous 
(F814W$<$21.5) red ((F606W - F814W)$ >$1.0) galaxies that also have
morphologies typical of early types (i.e., excluding galaxies with
clear disk-like structure).  As a sanity check, all three of the
lensing galaxies in this field satisfy the above criteria.  The
magnitude, color, and morphology cuts yield a conservative estimate of
14 potential lenses.  To test the validity of our cuts, we extended
them to one magnitude fainter and one magnitude bluer.  Of the 30
additional galaxies that were thus included, only two satisfied the
morphology criterion.  Therefore, we assume that our cuts have located
the vast majority of the luminous ellipticals in the field.  The
resulting total cross-section for lensing in this field is $\sim$16
arcsec$^2$.  This cross-section must be compared to the density of
faint blue galaxies. The completeness limit for the F606W imaging that
is presented in this paper is F606W$\sim$26.5\footnote{Here the
magnitudes are in the AB system, in order to compare to the results
of \citet{faintacsgals}.}.  Assuming a typical magnification factor
from strong lensing of a few, which is what we find from the lens
models, thus requires a knowledge of the integrated number density of
galaxies with F606W$<$28.  We use the corrected number counts
of \citet{faintacsgals} to obtain an integrated number density of
$\sim 2 \times 10^6$ galaxies per square degree or $\sim$0.2 galaxies
per square arcsecond.  The combination of this surface density with
the approximate lensing cross-section in the field yields an {\em a
posteriori} expectation value of $3.2 \pm 1.7$ lenses, given our
assumptions.  Therefore, it appears quite likely that two additional
lenses would be found in these images.

We conclude that an efficient method to search for new gravitational
lenses is to obtain deep images of the fields surrounding known strong
lenses.  This method takes advantage of the enhancement of the
line-of-sight number densities of both the potential lenses and
potential sources. We note that this is more an effect of biased
galaxy formation than a bias due to the enhancement of the lensing
cross-section by the group environment, which is secondary effect to
lensing statistics \citep[e.g.,][]{keeton_group, 1608groupdisc}.  
We also note that the results presented here suggest that clustering
of sources and lenses should be taken into account when comparing
observed and predicted lensing statistics in order, e.g., to place
limits on cosmological parameters.


\acknowledgments 

These observations would not have been possible without the expertise
and dedication of the staffs of the Palomar and Keck observatories.
We especially thank Karl Dunscombe, Grant Hill, Jean
Mueller, Gary Puniwai, Kevin Rykoski, Gabrelle Saurage, and
Skip Staples.
CDF and JPM acknowledge support under HST program \#GO-10158.  Support
for program \#GO-10158 was provided by NASA through a grant from the
Space Telescope Science Institute, which is operated by the
Association of Universities for Research in Astronomy, Inc., under
NASA contract NAS 5-26555.
This work is supported in part by the European Community's Sixth
Framework Marie Curie Research Training Network Programme, Contract
No.  MRTN-CT-2004-505183 `ANGLES'.
Based in part on observations made with the NASA/ESA Hubble Space
Telescope, obtained at the Space Telescope Science Institute, which is
operated by the Association of Universities for Research in Astronomy,
Inc., under NASA contract NAS 5-26555. These observations are
associated with program
\#GO-10158.
Some of the data presented herein were obtained at the W. M. Keck
Observatory, which is operated as a scientific partnership among the
California Institute of Technology, the University of California, and
the National Aeronautics and Space Administration. The Observatory was
made possible by the generous financial support of the W.M. Keck
Foundation.  The authors wish to recognize and acknowledge the very
significant cultural role and reverence that the summit of Mauna Kea
has always had within the indigenous Hawaiian community.  We are most
fortunate to have the opportunity to conduct observations from this
mountain.
Based on observations obtained at the Gemini Observatory, which is 
operated by the Association of Universities for Research in Astronomy, 
Inc., under a  cooperative agreement with the NSF on behalf of the Gemini 
partnership: the National Science Foundation (United States), the Particle 
Physics and Astronomy Research Council (United Kingdom), the
National Research Council (Canada), CONICYT (Chile), the Australian 
Research Council (Australia), CNPq (Brazil) and CONICET (Argentina).
%

\ifsubmode
  \newpage
\fi

%
%
%
%
%

\end{document}